# SplitArchitecture: SDN for the carrier domain

Authors: Wolfgang John, András Kern, Mario Kind, Pontus Sköldström, Dimitri Staessens, Hagen Woesner

## Abstract:

The concept of software-defined networking (SDN) has emerged as a way to address numerous challenges with traditional network architectures by decoupling network control and forwarding. So far, the use cases for SDN mainly targeted data-center applications. This paper considers SDN for network carriers, facing operation of large-scale networks with multi-million customers, multiple technologies and high availability demands. With specific carrier-grade features such as scalability, reliability, flexibility, modularity and virtualization in mind, the SPARC EU project has developed the SPARC *SplitArchitecture* concept. The *SplitArchitecture* control plane allows hierarchical layering of several control plane layers which can be flexibly mapped to data plane layers. Between control layers open interfaces are used. Furthermore, *SplitArchitecture* proposes an additional split of forwarding and processing functions in data path elements, enabling switch based OAM functionality and handling of tunneling techniques. The *SplitArchitecture* concept is evaluated in a prototype demonstrating a SDN version a Broadband Residential Access Server (BRAS): the *floating BRAS.* The *floating BRAS* allows creation of residential internet access services with dynamically managed BRAS instances. The demo is based on a controller enabling protected MPLS transport connections spanning SDN-controlled aggregation and IP/MPLS-based core domains. The prototype showcases how *SplitArchitecture* enables virtualization of service nodes in an SDN-controlled network, leading to increased flexibility in configuration and placement of service creation functions. Overall, the results show that it is technically and economically beneficial to apply SDN, and specifically the *SplitArchitecture* concept, to the carrier domain.

## 1 Intro

In the last decades the Internet has changed our lives. At the same time, its usage and hence the requirements on the underlying infrastructure and technologies are also constantly changing. It is widely accepted that numerous technical challenges e.g. in areas like reliability, scalability or security, are the result of this evolution. Unfortunately, these challenges seem to demand more comprehensive efforts than simple patches, like adding new data- or control plane protocols. Manifold research activities have been started to develop appropriate solutions, such as the Future Internet [1] or the Clean Slate Internet program [2]. A major result of the latter is the idea of splitting the forwarding layer (i.e. the data plane) and the control layer, which enables independent evolution of the control and forwarding solutions. Together with the split, an open interface between the layers is proposed. The most prominent protocol implementing such an interface - OpenFlow - was originally developed at Stanford University and is now marshalled by the Open Networking Foundation (ONF) [3]. Utilizing open interfaces like OpenFlow to configure the network led to the wider concept of Software Defined Networking (SDN), where the network operations are determined by control programs running at a logically centralized controller (sometimes referred to as network operating system, see Figure 1). Much research focuses on the interface between control layer and data plane. Research has also shown how SDN can facilitate network abstractions, similar to the role of hypervisors used in server virtualisation [4].

The usage scenarios of OpenFlow and SDN have so far primarily focused on research (e.g., [5], [6]), data centre and enterprise network environments. Private wide area networks have been included in the scope, but public telecommunication / carrier networks are still studying SDN [7][8]. We argue that carrier networks could also greatly benefit from the advantages of SDN, i.e. simplifying introduction of new features and facilitating operations [3]. The dominant advantage is the improved flexibility and modularity provided by the split of control layer and data plane. Today, carriers are struggling with the maturity and standardisation of packet technologies like various flavours of Ethernet, MPLS or IP, requiring among other things thorough testing activities to ensure the compatibility with devices as well as the interoperability of equipment. As a result, any modification of the traditional technology stack starts processes which demand substantial resources in terms of time and manpower. With SDN, network elements or even platforms could be partly reused while adding new features to the control layer, with the promise to reduce investments significantly.

The reasons why carrier networks have not been a focus of SDN research can be found in the specific constraints of the carriers' environments and the related requirements:

- Millions of end-customers and endpoints as well as hundreds of peering points, and network structures that require *scalability*
- Untrusted environment with demand for Authorisation, Authentication and Accounting with strong Service Level Agreements requiring special extensions in the area of *reliability*
- Support for multiple network providers (wholesale services) sets requirements for a high levels of isolation in *virtualisation* solutions
- Sophisticated administration and management support is required for operational cost optimisation
- Legacy platforms and customer equipment (e.g. ATM, MPLS, different Ethernet flavours, SDH, etc.) with available migration options need to be taken into account and require *flexibility* and *modularity*

The SPARC[1] project aimed at applying SDN principles based on the OpenFlow protocol for developing a control solution which provides the promised flexibility in carrier grade operation of networks and the desired benefits in capital- and operational expenditures. SPARC was concentrating on the access/aggregation network domain (sometimes called metro network segment) of carrier networks which connects residential and business end-customers with a carriers' core network.

---

[1] FP7 SPARC (Split Architecture for carrier-grade wide-area networks) was a publicly funded research project by the European Commission under grant agreement number 258457.

# 2 SPARC *SplitArchitecture* concept

Applying SDN concepts in OpenFlow networks led to strongly centralized architectures, where a central entity, often called controller, implements all network control functions. The major functional elements commonly seen in such central entity are depicted on the left side of figure 1. We derived this generalized SDN model from existing SDN models currently seen in industry and academia [3][4]. The virtualization layer (network hypervisor) has a global view of the underlying physical topology and its resources. The network operation system (Network OS) constitutes the set of fundamental functions that must be provided (e.g. topology discovery and node repository handling). Control programs (e.g. different routing algorithms) are network-related functions, which are not part of the essential set of functions provided by the network operation system itself. These control programs provide APIs to other applications, e.g. business applications.

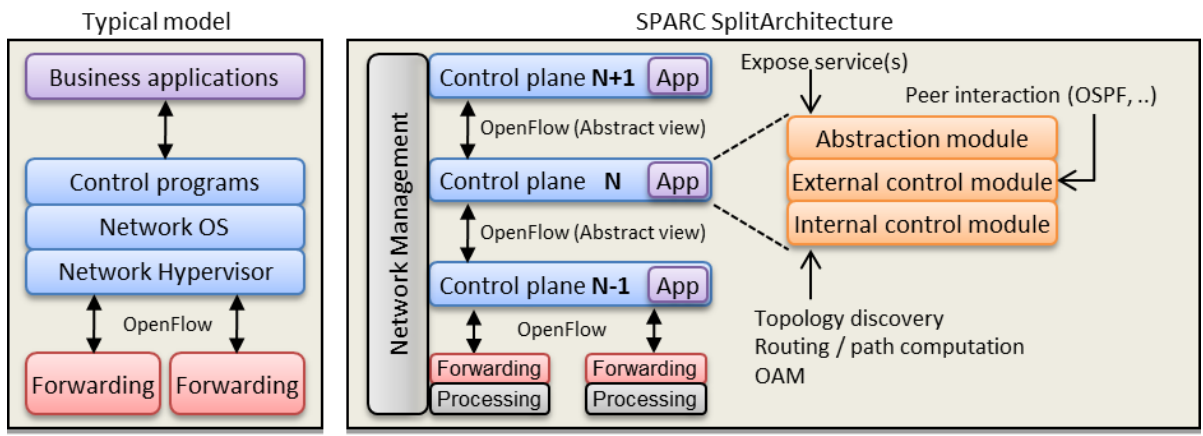

Figure 1: A generalized SDN model typically seen in literature (left box); the proposed SPARC *SplitArchitecture* (right box).

To achieve forwarding plane scalability, the connectivity structure commonly implements hierarchies, which also provides flexibility via standardized adaptation functions (e.g., pseudo wire emulation allows carrying Ethernet on top of MPLS). One possible way to reach corresponding *scalability* and *flexibility* also for the control layers is to allow the mapping of data plane layers to several control plane layers[2] (see right side of Figure 1). Basically, the functions of the network operating system, the hypervisor as well as the control programs (or applications) are implemented in each control plane layer. Several of these control layers, stacked upon one another, provide the ability to adjust and adapt the overall control plane to varying network scenarios and technologies by applying well-tested functional control blocks. This hierarchical controller architecture is what we refer to as *SplitArchitecture* [9].

Between these control layers, open and standardized interfaces are desired in order to improve the *flexibility* of the hierarchy. This enables operators to deploy parallel control planes with minimal interference with each other. Furthermore, it gives a network operator tight control over the level of detail of which data plane details are exposed to higher control planes or to third parties. It is preferred to use the same protocol to implement (part of) the interface towards higher layer control entities as the one that controls the data plane elements. One realization used in *SplitArchitecture*, is to apply the OpenFlow protocol not only between controllers and the switches but between controllers, too. Thus, each entity controlling data path elements in a lower plane emulates a single data plane entity toward higher control planes, one for each client. Essentially, we replace a separate, central network hypervisor by a basic set of filter functions in each control entity, providing a filtered, abstract network view to higher control planes. This function also provides the ability to share the physical network between well isolated *virtual* networks, e.g. for multi-tenant and multi-service scenarios (i.e. network *virtualization*). Currently, the ONF discusses the implementation of controller hierarchies for multi-domain and scalability issues as Intermediate Controller-Plane-Interface (I-CPI) [10]. The SPARC *SplitArchitecture* has inspired these ongoing discussions as one common solution for implementing ONF's I-CPI as well as the two interfaces Application- and Data-Controller-Plane-Interfaces (A-/D-CPI). SPARC *SplitArchitecture* would allow for implementation of one interface (i.e. an adopted OpenFlow variant) and reduce the complexity to the information models carried via these interfaces.

---

[2] As a corner case, the control plane can also be implemented as a single layer, which would correspond to a generalized SDN model.

We expect that similar functions are implemented at every control layer (depicted at very right part of Figure 1). Control functions are grouped into sub-blocks. First, an internal control block hosts the control logic providing connectivity services within the domain under control (functioning like a backplane for the controlled network domain). Second, an external control block contains the control logic which interacts with peers of the same control layer managing other domains. Finally, an abstraction module exposes an abstracted view of the network via the OpenFlow protocol in form of a logical switch with a set of ports.

Finally, the *SplitArchitecture* concept also includes a refinement of the data plane model, by further splitting the data plane switch into forwarding and processing functions, such as adding/removing headers, analyzing and performing OAM messages, etc. A main motivation for the introduction of processing capabilities in the data plane, which is somewhat against the original OpenFlow idea of "dumb" switches, is the stringent *availability* and *reliability* demands (e.g. less than 50 milliseconds restoration times). Another motivation is the need to handle protocols on the switches that requires processing capabilities beyond existing OpenFlow actions, for example many tunneling techniques commonly used in carrier networks (PWE, PPPoE, etc.).

# 3  The "Floating BRAS" case study

A current trend is that MPLS, which has been predominant in core networks for many years, extends further into the aggregation network. Covering all network segments with a common packet forwarding technology promises simpler network control. However, when the size of the IP/MPLS network segment grows from hundreds to tens of thousands of nodes, the simple IP/MPLS control plane deployments suffer scalability issues. Control plane architectures, like Seamless MPLS [11], handle these issues at the cost of additional configuration and increased control plane complexity at every network node.

In theory it is possible to instantiate service creation points at any node of the network on top of a unified MPLS based transport network. An example of common service creation points for residential Internet access would be Broadband Residential Access Servers (BRAS), in current networks typically located at the edge of the operators IP core network.

Our case study applies the *SplitArchitecture* concept to provide scalable transport control of an MPLS-based unified aggregation and core network. We raised the question how service creation can be harmonized with transport control, resulting in a flexible network control solution in-line with SDN principles. To answer this question, we consider residential Internet access service based on PPP.

The control solution must realize three major functions: establish and maintain MPLS transport services between any two nodes of the network (see Section 3.1); control the provided residential IP service (Section 3.2); and optimize service creation and transport services (Section 3.3). These three functional groups set different requirements. The deployment of MPLS transport services expects predefined control functions, such as topology discovery, connection setup and restoration. In practice, MPLS is the only forwarding technology to be supported. This aspect shows that for the transport services, flexibility is less important than scalability: the control functions need to be executed as fast as possible. In contrast, IP services are much more diverse, with differing characteristics and frequent changes. Thus SDN-aware service creation additionally calls for *flexibility* and *extensibility*. Obviously, obeying the different requirements to realize the different control function blocks as shown in Figure 2 will lead to a mixed control architecture.

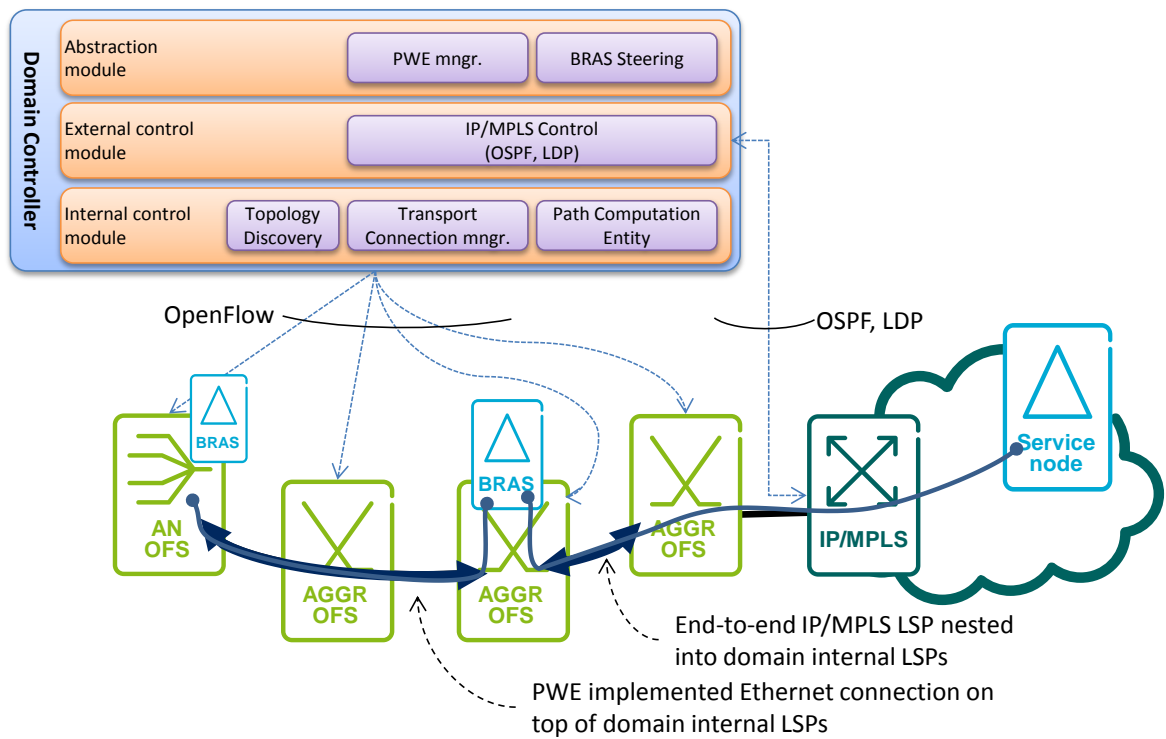

Figure 2: Implementing SPARC *SplitArchitecture* - key functional elements, interfaces and protocols

## 3.1 MPLS transport control

In our case study, the MPLS transport control plane is heterogeneous: in the core, the distributed IP/MPLS control plane is kept; In the aggregation, together with the introduction of MPLS forwarding, centralized aggregation domain controllers are deployed that supervise the Access Nodes and Aggregation Switches (AN OFS and AGGR OFS, respectively) forming the aggregation network. This results in a mixed control plane formed of SDN domain controllers and IP/MPLS routers. Using IP/MPLS control protocols (OSFP, LDP, MP-BGP) between the controller and the IP/MPLS core eliminates any need of updating the IP/MPLS routers' control plane. For enhanced scalability purposes, the whole aggregation domain is represented as a single IP/MPLS router. The external control module, besides running the protocol procedures, carries out the essential virtualization functions. Hiding the topology and configuration details from the core control plane allows reacting locally on network triggers, such as failure without notifying the core. To support all above features while fulfilling the scalability requirement, our MPLS transport controller [12] is realized through a single control plane layer of the SPARC *SplitArchitecture* and applies the three control blocks (see Figure 2):

The internal control module maintains the MPLS connections in the aggregation domain. As part of this task, this control module performs topology discovery. Besides discovering the topology, this function is able to detect any topology changes, in order to trigger other control functions that react on link failures. A transport connection manager function is responsible for keeping track of the established aggregation-domain internal LSPs. This module installs new LSP upon requests of other control modules and is able to reconfigure the already installed LSPs in case of network failures. This function comprises of a path computation entity, which calculates different path types: multipoint-to-point paths for carrying all MPLS traffic to certain destination over one LSP (referred to as merging tree); and a pair of point-to-point paths to support 1 to 1 path protection and multicast trees. Once a new path is calculated, the control blocks update the switch configuration using OpenFlow. When the topology discovery detects link failures it triggers path computation to recalculate the merging trees.

The external control module participates in signaling end-to-end MPLS LSPs, which span over multiple domains, e.g., starting in one aggregation domain, running through the core and terminating in another aggregation domain. As the core network is assumed to be IP/MPLS, this control block interacts with peering routers using standard IP/MPLS protocols. When the control module initiates or receives signaling messages of an end-to-end LSP, it initiates the configuration of the local switches. The instantiation of a new end-to-end LSP is performed as follows: First, the control module determines the domain ingress and egress switches, at which the end-to-end LSP enters and leaves the control domain. If the end-to-end LSP starts or terminates at the control domain the control module also determines the LSP ingress or egress nodes. Then, it requests the internal control module to form a co-routed bidirectional domain-internal LSP implemented by a pair of unidirectional merging LSPs between these nodes. Finally the control module configures ingress and egress nodes to nest the end-to-end LSP into the server LSPs. The head-end and tail-end nodes are configured to be act as LSP endpoints.

On top of these LSPs, the abstraction module of the MPLS transport controller implements MPLS transport services using pseudo wires emulation (PWE). The necessary control of PWs is implemented by the PWE Manager element sitting on the top of the external and internal control modules. This element will configure the PW endpoints, and requests the other control modules to create the necessary MPLS LSPs between the PW capable nodes. However, the service creation element does not have direct control of the remote PW capable nodes when they reside in other control domains. There are several ways of coordinating the PW configuration, e.g., using targeted LDP sessions or direct configuration of the remote PW endpoint by the service creation element.

One important aspect of carrier grade requirements is the ability to define stringent restoration thresholds (e.g. 50 milliseconds). This requirement challenges the hierarchical control plane design, as all control functions are moved to a logically centralized, distant controller which must detect link failures and reconfigure affected flows within the given time constraints. Experiments conducted within the SPARC project confirm that controller based restoration is indeed insufficient to provide high availability in carrier grade networks, since it scales linearly with the number of affected flows in the network [13]. Switch based protection, on the other hand, would scale well and would thus be able to fulfill the timing requirements. In order to realize protection, OpenFlow 1.1 introduced *fast failover* groups, a mechanism to perform protection switching within a data path element. However, due to the lack of a path continuity check mechanism, fast failover group entries can only rely on errors detected on the local port. While this mechanism is well suited for link protection, solutions that reserve less backup capacity, like shared backup path protection, cannot be implemented without active involvement of the controller.

We have extended the switches with continuity checking capabilities, realized by tight integration of monitoring packet generation and handling into the switch pipeline [14]: template monitoring packets are generated and enter the switch via logical ports. As the template packets traverse the packet processing pipelines they are completed (i.e. header fields are filled in), as specified by actions. Complete monitoring packets are injected into a data flow, from where the egress node can filter them and send them for processing in its local OAM processing entity. If a failure is detected the processing entry triggers local protection switching and notifies the controller of the event. As a result, our switch based protection mechanism offloads the controller by placing continuity checking and protection switching in the switches.

## 3.2   Creating residential Internet access in SDN fashion

For residential access, Internet connectivity is typically the first established service. The customer's terminal or home gateway is attached to the Internet via a Broadband Residential Access Server (BRAS). One implementation of a direct logical link between the customer premises equipment and the BRAS is the point-to-point protocol (PPP). The adaptation to link layer technology used in the operator's aggregation network (e.g., ATM or Ethernet) also must be solved (using PPPoA or PPPoE). In the PPPoE case a discovery and negotiation phase is added to service creation procedures where the client can select from potentially different BRAS (for resiliency) or service providers. Other options to implement residential Internet access service also exist, such as DHCP based control and authentication. Here the customer circuits are identified with a pair of customer and service VLAN identifiers and the customer IP datagrams directly encapsulated into such double-tagged Ethernet frames. Though this is just a simple service example, the efficient realization and management of the different implementation calls for applying the SDN principles.

In our case study, we developed a standalone software module in charge of basic BRAS functions, i.e. involving PPPoE discovery, PPP session control, management of customer IP addresses, and routing of customer packets. Control and forwarding planes are split and the BRAS module is connected to the main forwarding function, provisioned by the transport controller (see Figure 3). The forwarding functions of the data path elements are using node-internal virtual ports. The module also includes PPP and PPPoE processing functions. As a result, the split BRAS implementation introduced a fine split of control layers according to the hierarchical *SplitArchitecture* control plane, with dedicated control layer entities for the Ethernet and the IP switching layers, respectively, and adaptation functions implementing encapsulation functions (e.g., PPP).

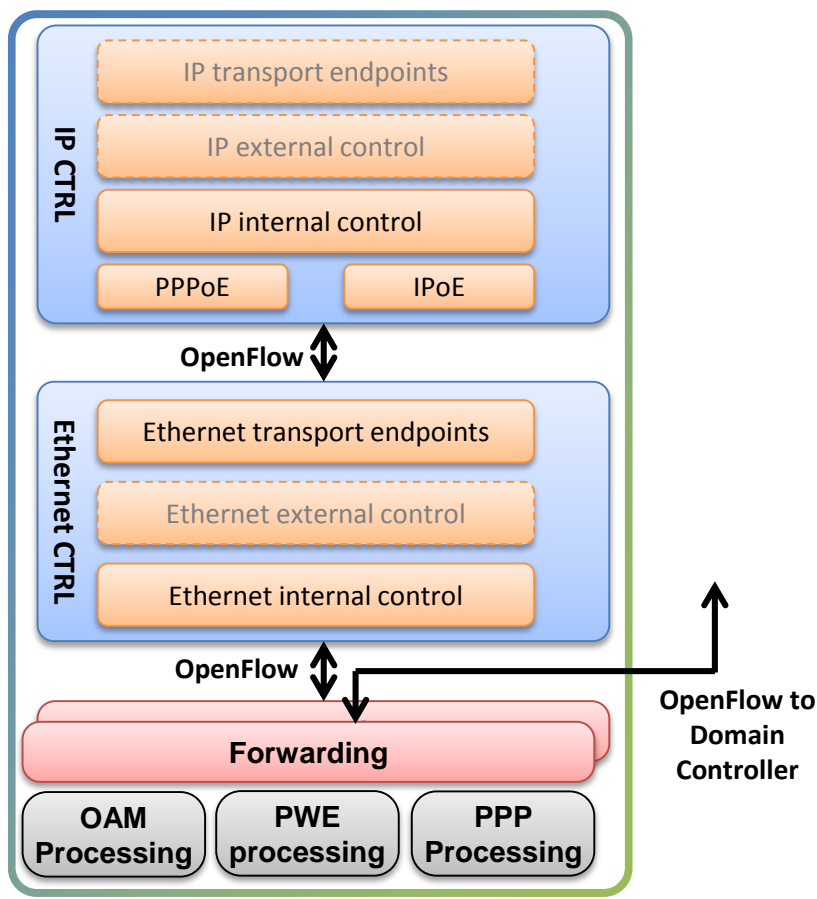

Figure 3: Adopting hierarchical control for implementing split BRAS

The Ethernet control layer provides a virtual node view to the IP controller by substituting the switch internal virtual ports with logical ports representing Ethernet flow endpoints. The IP controller performs routing decision in its internal control block based on static IP routes installed when a new customer is attached to the BRAS. Between the IP and the Ethernet controllers, two types of encapsulation are essential. Towards the users IP frames are encapsulated using PPPoE. This encapsulation together with the associated control procedures (PPP link control and monitoring messages) are implemented as a logical port. The other encapsulation is IP over Ethernet, which is also modeled as a logical port. In the implementation, the BRAS module does not peer with other control entities, thus no external control blocks were implemented for either Ethernet or IP controllers. Since the BRAS does not originate or terminate IP traffic it does not provide IP transport endpoints.

The OpenFlow protocol is used for two purposes: passing data packets between the control layers and for a client layer to send flow configuration rules to its server. For instance, after a routing decision, the IP controller formulates a bypass flow rule encoding this decision. However, this rule is incomplete in the sense that it does not have instructions for handling other fields, e.g. Ethernet header fields. As the flow rule traverses toward the fast path, the logical ports and the Ethernet controller add these instructions. When the flow rule reaches the switch it is installed into a flow table and subsequent packets matching these rules will no longer be sent to the controller, they are processed and forwarded by the fast path. This combination of Ethernet and IP controllers with the data plane functions leads to an extremely configurable and flexible realization of an IP router: introducing a new service translates to adding an additional controller to the controller chain.

## 3.3  Service Creation and Optimization

With the BRAS module it becomes possible to instantiate BRAS functions at any aggregation switch able to support the processing functions required by the split BRAS module [15].

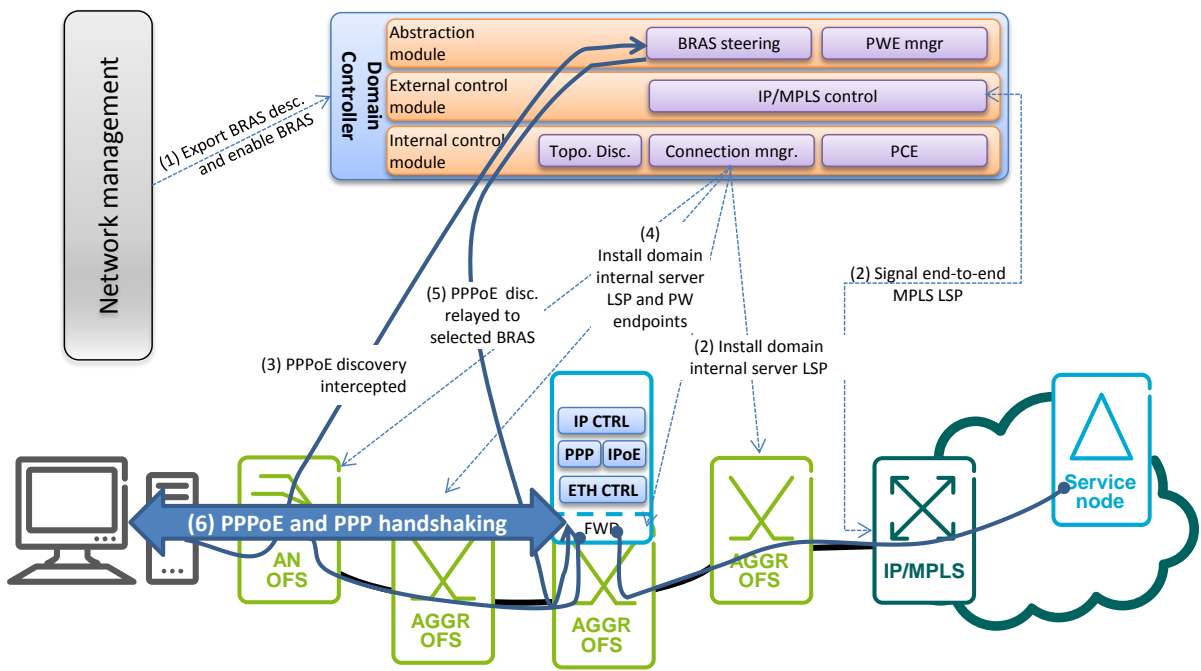

Figure 4: Creating residential Internet access service with *SplitArchitecture*

As depicted in Figure 4, the network management system (NMS) is in charge of starting and configuring the BRAS modules at selected aggregation switches. The NMS also exports BRAS descriptors to the centralized control plane, which has a BRAS Steering Module implementing the following service creation control functions.

The NMS dynamically defines which of the instantiated BRAS modules can be used. When a BRAS module is enabled, the BRAS Steering Module creates bidirectional connectivity between the BRAS hosting node and the next-hop service node, typically a gateway connected to the Internet. Since these gateways can sit at the far end of the network, the BRAS Steering Module invokes the external control module to create end-to-end IP/MPLS tunnels to establish bidirectional connectivity. When a BRAS module is disabled this connectivity might be torn down.

In order to get Internet access the end user first initiates PPP session discovery. This message enters the aggregation network at the access node, which relays it to the centralized controller. In the controller the BRAS Steering Module may authenticate the packet and selects one of the enabled BRAS modules. The Steering Module then establishes Ethernet connectivity between the customer facing access node and the node hosting a BRAS module. In case of MPLS based transport this means instantiation of pseudo wires on the top of MPLS tunnels. Once the transport connectivity service is established, i.e. all tunnels are established, the BRAS Steering Module passes the PPP discovery packet to the selected BRAS module. The BRAS module then continues the PPP handshake using the direct transport tunnels established to the user.

Selection of a BRAS module, in our implementation, is based on pre-assigned priorities. This policy can be extended to any arbitrary policy and may e.g. be based on statistical information provided by the BRAS modules.

## 4 Conclusions

In this paper we discussed the application of the SDN concept to the network operator domain, with the goal of improving network design and operation in large-scale networks with multi-million customers, heterogenous technologies and high availability demands. To fulfill these requirements we defined scalability, flexibility, modularity and virtualization as key control plane guidelines. Based on these guidelines we developed the SPARC *SplitArchitecture* concept.

As in hierarchically structured connectivity at the forwarding plane, the *SplitArchitecture* control plane also allows hierarchical layering of several control plane layers which can be flexibly mapped to data plane layers. Thus each entity that controls data path elements in a lower plane emulates a single data plane entity toward higher control planes. Between control layers open interfaces are used - in our case OpenFlow. This control plane architecture achieves outstanding flexibility by plugging various control modules together to form a hierarchical control plane. Providing virtual data plane entities to different clients brings the ability to share the network resources, e.g. for multi-tenant, multi-service scenarios through network virtualization. Finally, *SplitArchitecture* also proposed an additional split of forwarding and processing functions of data path elements.

Processing capabilities in the data plane enable switch based OAM functionality, which is not only crucial to meet operators stringent reliability demands, but also allows handling of tunneling techniques commonly used in carrier networks (PWE, PPPoE, etc.).

To evaluate this concept, we applied the SPARC *SplitArchitecture* to the *floating BRAS* use case, which allows creation of residential Internet access services with dynamically managed BRAS instances. The demonstrator was done in two stages: In the first stage, we developed a controller enabling carrier-grade MPLS transport connections by applying *SplitArchitecture*. The resulting controller is able to provide protected MPLS LSPs, using the OpenFlow protocol over a set of aggregation nodes, participating via IP/MPLS to create end-to-end connection spanning both aggregation and core domains. In the second stage, the split BRAS module was integrated in the demonstrator, resulting in the successful floating BRAS demo. The floating BRAS prototype showcases how *SplitArchitecture* enables virtualization of service nodes in an SDN controlled network, leading to increased flexibility in configuration and placement of service creation functions.

Regarding business aspects, the SPARC project also provides a techno-economic analysis on the applicability of SDN in operator networks (e.g. mobile backhaul networks) [16]. The results confirm that SDN can provide substantial cost reductions in terms of capital (CAPEX) and operational expenditures (OPEX), even for carrier networks. Hence, the conclusion of the SPARC project is that it is both technically feasible and economically beneficial to apply SDN, and specifically SPARC *SplitArchitecture*, concepts to the carrier domain.

## Acknowledgements

This work was funded by the European Commission under theFP7 ICT research project SPARC (258457). The authors would like to thank all SPARC partners for discussions and comments.

## Author Biographies

**Wolfgang John** studied computer science and engineering in Salzburg, Halmstad, and Göteborg. He received his PhD degree from Chalmers University of Technology (2010), researching on measurement and analysis of Internet traffic. Since 2011 he is with Ericsson Research in Sweden, working on network management for SDN, NFV and Cloud environments, including participation in FP7 ICT projects SPARC and UNIFY. Wolfgang authored multiple journal articles, conference papers, and patent applications, and is involved with ONF standardization.

**András Kern** received his M.Sc. (2003) and Ph.D (2008) degrees from Budapesti Műszaki és Gazdaságtudományi Egyetem (BME). Since 2007, he is with Ericsson Research, Hungary. He is the author of conference and journal papers and holds several patents. He is active in the software defined networking (SDN) area. He was a technical leader in the SPARC EU project focusing on OpenFlow implementations and prototyping in the SPARC project.

**Mario Kind** works as Professional R&D Engineer for Telekom Innovation Laboratories, the research department of Deutsche Telekom AG. During the last eight years in research he was evaluating the business, technology, and service trends in the Internet, telecommunication, and broadcast industry and consults on the impact on the future of broadband access networks. Currently, he is challenged by the new possibilities through SDN topics especially by participation in research projects FP7 ICT SPARC and UNIFY.

**Pontus Sköldström** received an MScE from the Royal Institute of Technology (KTH) in 2008. Since then he is with ACREO, researching and developing transport-, access- and home networks. In 2011 he began his PhD studies at KTH. In FP7 ICT projects ALPHA, SPARC and UNIFY, Pontus has an ongoing focus on future network architectures, including GMPLS control planes, OAM and network virtualization for SDN, and orchestration of NFV environments. He is involved in IETF standardization.

**Dimitri Staessens** received his M.Sc. Degree in numerical computer science in 2004 from Ghent University, Belgium. Since 2005 he has been working at the "Internet Based Communications Networks and Services group" and finished a PhD on survivability of optical networks in 2012. His work is focusing on control and management of networks, Software Defined Networking and future network architectures, and performed in EC ICT projects such as NOBEL, BONE, SPARC, CITYFLOW, IRATI and PRISTINE.

**Hagen Woesner** studied computer science in Chemnitz and Berlin, receiving both Master and PhD degree from TU Berlin. His research interests range from wireless to optical packet networks. After a post-doc at CREATE-NET (2006), he was leading the IT/Infrastructure group of EICT, coordinating the FP7 ICT project OFELIA and contributing to SPARC. He is chairing the recurring European Workshop on SDN (EWSDN), and leads BISDN GmbH as co-founder.